\documentclass[aps, superscriptaddress, twocolumn,prl,floatfix]{revtex4}
\usepackage{amssymb}
\usepackage{amsmath}
\usepackage{graphicx}

\def\SiOx{\mathrm{SiO}_2}
\def\den{n_{s}}

\def\Vbg{V_{\mathrm{G}}}
\def\Vt{V_{\mathrm{T}}}
\def\It{I}
\def\gt{g}
\def\dgt{\delta g}

\begin{document}
\title{Tunneling Spectroscopy of Graphene-Boron Nitride Heterostructures}
\author{F. Amet}
\affiliation{Department of Applied Physics, Stanford University, Stanford, CA 94305, USA}
\author{J. R. Williams}
\affiliation{Department of Physics, Stanford University, Stanford, CA 94305, USA}
\author{A. G. F. Garcia}
\affiliation{Department of Physics, Stanford University, Stanford, CA 94305, USA}
\author{M. Yankowitz}
\affiliation{Department of Physics, Stanford University, Stanford, CA 94305, USA}
\author{K.Watanabe}
\affiliation{Advanced Materials Laboratory, National Institute for Materials Science, 1-1 Namiki, Tsukuba, 305-0044, Japan}
\author{T.Taniguchi}
\affiliation{Advanced Materials Laboratory, National Institute for Materials Science, 1-1 Namiki, Tsukuba, 305-0044, Japan}
\author{D. Goldhaber-Gordon}
\affiliation{Department of Physics, Stanford University, Stanford, CA 94305, USA}

\date{\today}

\begin{abstract}
We report on the fabrication and measurement of a graphene tunnel junction using hexagonal-boron nitride as a tunnel barrier between graphene and a metal gate.  The tunneling behavior into graphene is altered by the interactions with phonons and the presence of disorder. We extract properties of graphene and observe multiple phonon-enhanced tunneling thresholds. Finally, differences in the measured properties of two devices are used to shed light on mutually-contrasting previous results of scanning tunneling microscopy in graphene.

\end{abstract}
\maketitle

The probability for an electron to tunnel through a very thin potential barrier depends strongly on the number of available states in the target material at the same energy. Tunneling spectroscopy uses this principle to directly probe the density of states of complex materials whose electronic properties are still poorly understood~\cite{TunnelReview}. Furthermore, the phonon energy spectrum of a material can be determined by measuring the second derivative of the tunneling current~\cite{TunnelReview}. Thus, a tunneling measurement can obtain information on both the electronic and phononic properties of a material. 

In particular, tunneling experiments have been performed on graphene,  a carbon monolayer with a honeycomb lattice where electrons are confined in two dimensions. In this material, electrons behave like relativistic particles obeying the Dirac equation,  leading to a variety of exciting electronic behaviors~\cite{CastroNeto09}. In addition to probing the density of states, prior tunneling spectroscopy measurements on graphene revealed features associated with interactions between electrons, phonons and disorder~\cite{Zhang08, Zhang09, Jung11,Malec11}. At low carrier densities, charge inhomogeneities modify electron tunneling in scanning tunneling microscopy (STM)~\cite{Zhang09, Jung11}. However, STM experiments have exhibited conflicting results. While some STM measurements have shown a strong suppression of tunneling at low energies~\cite{Zhang08, Brar07}, attributed to interactions between electrons and K-point phonons in graphene~\cite{Wehling08}, other experiments have not observed this effect, whether for graphene on $\SiOx$~\cite{Jung11, Li09} or graphene on hexagonal-boron nitride (h-BN)~\cite{Xue11}.

In this Letter, we complement the existing tunneling experiments by fabricating graphene tunnel junctions using h-BN as a tunneling barrier. We focus on two devices, referred to below as A and B. The devices were fabricated by the same process, but as we will see below they exhibit different tunnel behavior: one dominated by the density of states in graphene and one dominated by the energy dependence of the transmission probability. Each resembles a different set of prior STM experiments. Properties of graphene extracted from the tunnel measurements, such as disorder-induced charge puddle size and Fermi velocity, agree well with previously reported results obtained via STM. In addition, we observe a rich electron-phonon interaction structure in the inelastic tunneling current, allowing for the identification of resonances near the phonon energies at the K and M-point in graphene as well as two low energy resonances that we attribute to Van Hove singularities in the h-BN phonon density of states. Finally, the energy dependence of the tunnel transmission is strongly influenced by the microscopic details of the device geometry, shedding light on previously mutually-contrasting results in STM experiments. 

\begin{figure}[t!]
\center \label{fig1}
\includegraphics[width=3 in]{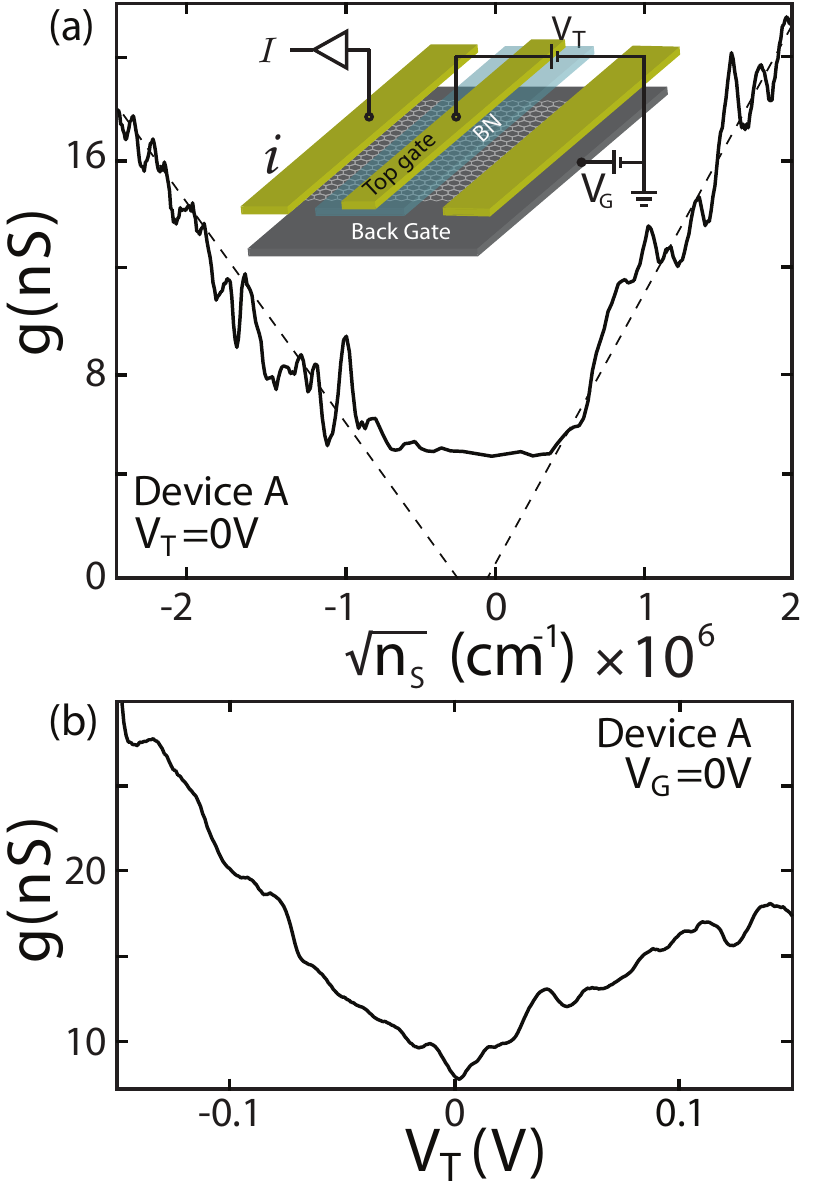}
\caption{\footnotesize{(a) (inset) schematic of the device structure used. An atomically-thin sheet of h-BN separates the top gate from graphene. An adjacent lead $i$ is used to extract the current $\It$ that travels from the top gate (biased by $\Vt$) to graphene. A voltage $\Vbg$ is applied on the back gate to control the sheet density ($\den$) in graphene. (main) The tunnel conductance $\gt$ as a function of $\sqrt{\den}$ measured at $\Vt$=0 and temperature of 4K. Dashed lines are guides to the eye. (b) $\gt$ measured as a function of $\Vt$ for $\Vbg$=0.}}
\end{figure}

The use of h-BN as an insulating substrate has allowed for high-mobility graphene devices to be fabricated~\cite{Dean10}. We exfoliate h-BN and graphene on separate degenerately-doped silicon wafer pieces capped with thermally-grown, 300\,nm-thick $\SiOx$. The thicknesses of the h-BN flakes used are measured with AFM to be $\approx$ 2\,nm. After annealing both flakes in Ar and H$_2$ at 350\,$^{\circ}$C for 4 hours each with flow rates of 500 sccm, the h-BN flakes are coated with a thick PMMA layer and the underlying $\SiOx$ is etched away with a solution of potassium hydroxide, which lifts off the PMMA with the h-BN attached to it. The PMMA-h-BN membrane is then attached to a micro-manipulator arm, similar to Ref.~\cite{Dean10}, and transferred on top of the graphene sheet. The PMMA is then dissolved away in acetone. Standard e-beam lithography techniques are employed to electrically contact the graphene with 10\,nm/50\,nm of Ti/Au and to add a top gate, which serves as a tunnel contact, on top of the h-BN flake [see inset of Fig. 1(a) for a schematic of the completed device].

The differential tunnel conductance $\gt \propto \partial \It/\partial \Vt$, where $\It$ is the tunnel current, is obtained as a function of the top-gate voltage $\Vt$ and backgate voltage $\Vbg$ using a lock-in measurement with an excitation voltage of 1\,mV at 13\,Hz.  $\Vbg$, applied to the degenerately-doped Si wafer, controls the average density of electrons $\den$ in the sheet of graphene. $\Vt$ both modulates $\den$ directly beneath the top gate and allows for the DC voltage bias dependence of the tunneling process to be probed. All experiments are performed at a temperature of 4.2\,K. In-plane transport measurements on the devices show a charge-neutrality point (CNP) of 18(-22)\,V for device A(B).

A plot of $\gt$ as a function of $\sqrt{\den}$ for device A is shown in Fig. 1(a); here, $\den$ is controlled by $\Vbg$, with $\Vt$ grounded. The tunnel conductance is proportional to $\sqrt{\den}$ away from the CNP, as shown by the dashed lines, directly reflecting the V-shaped behavior of the density of states. The variation of the density of states due to the metallic top gate is negligible within the range of $\Vt$~\cite{TunnelReview}, and $\It$ is expressed as
\begin{eqnarray}
\It(\Vt) \propto \int^{0}_{-e\Vt}\rho(E_{F}+\epsilon)T(\epsilon,e\Vt)d\epsilon
\end{eqnarray}
where $\rho(E)$ is the density of states in graphene, $E_F$ is the Fermi energy and $T(\epsilon,e\Vt)$ is  the energy and bias dependent electron tunneling transmission probability, which decays exponentially with distance and depends on the device geometry~\cite{SuppInfo}. $\rho(E)$ and $E_{F}$ are given by:
\begin{subequations} \label{allequations}
\begin{eqnarray}
 \rho(E)=\frac{2\vert E\vert}{\pi(\hbar v_{F})^{2}} \label{equationa} \\
 E_{F}=\hbar v_{F}k_{F}=\hbar v_{F}\sqrt{\pi \den},\label{equationb} 
\end{eqnarray}
\end{subequations}
where $v_F$ is the Fermi velocity of electrons.  At $\Vt$=0,  the tunnel transmission is constant and $\gt$ depends only on the density of states: $\gt(\Vt=0,\Vbg)\propto e\rho(E_{F}(\Vbg)) \propto \sqrt{\den}$, as observed in Fig. 1(a).  However, the density of states does not reach zero when $\den$ vanishes but instead flattens out when $\den \sim$ $5\times 10^{11}cm^{-2}$, due to charged-impurity disorder on the graphene flake~\cite{DasSarma11}. When $\den$ becomes as low as the density of impurities, p or n-type charge puddles form around the impurities~\cite{Martin08}.  The slopes of $\gt$ on the p-type and n-type regions differ by 20\%, indicating that $v_F$ is 9\% lower for electrons than for holes. This asymmetry is consistent with the presence of negatively-charged adsorbates on the sheet of graphene~\cite{Robinson08, Lohmann09}.

The $\Vt$-bias dependence of $\gt$ for $\Vbg$=0 is shown in Fig. 1(b). Results for device A show a smooth evolution of $\gt$ with $\Vt$, indicating no or relatively small variation in $T$($\Vt$), similar to Refs.~\cite{Jung11, Li09, Xue11}. In particular, there is no exponential suppression of $\gt$ near zero-bias, as was observed in Refs.~\cite{Zhang08, Brar07, Malec11}.

$\gt(\Vt,\Vbg)$ is moderately suppressed around $\Vt$=0 and along a diagonal region [Fig. 2(a), outlined by white dashed lines]. This region corresponds to gate voltages such that $\den$ is close to the CNP~\cite{Huard07, Williams07}, where the density of states is minimal. The tunnel probe also gates the graphene sheet and the relative back(top) gate capacitances $C_{\mathrm{G}}$($C_{\mathrm{T}}$) can be determined by fitting $C_{\mathrm{T}}\Vt+C_{\mathrm{G}}\Vbg$= constant. The slope of this line gives the capacitance ratio $C_{\mathrm{T}}/C_{\mathrm{G}}\approx 72$, which is lower than the ratio of 150 that one expects from a simple parallel plate capacitor model~\cite{SuppInfo}. 

\begin{figure}[!]
\center \label{fig2}
\includegraphics[width=3 in]{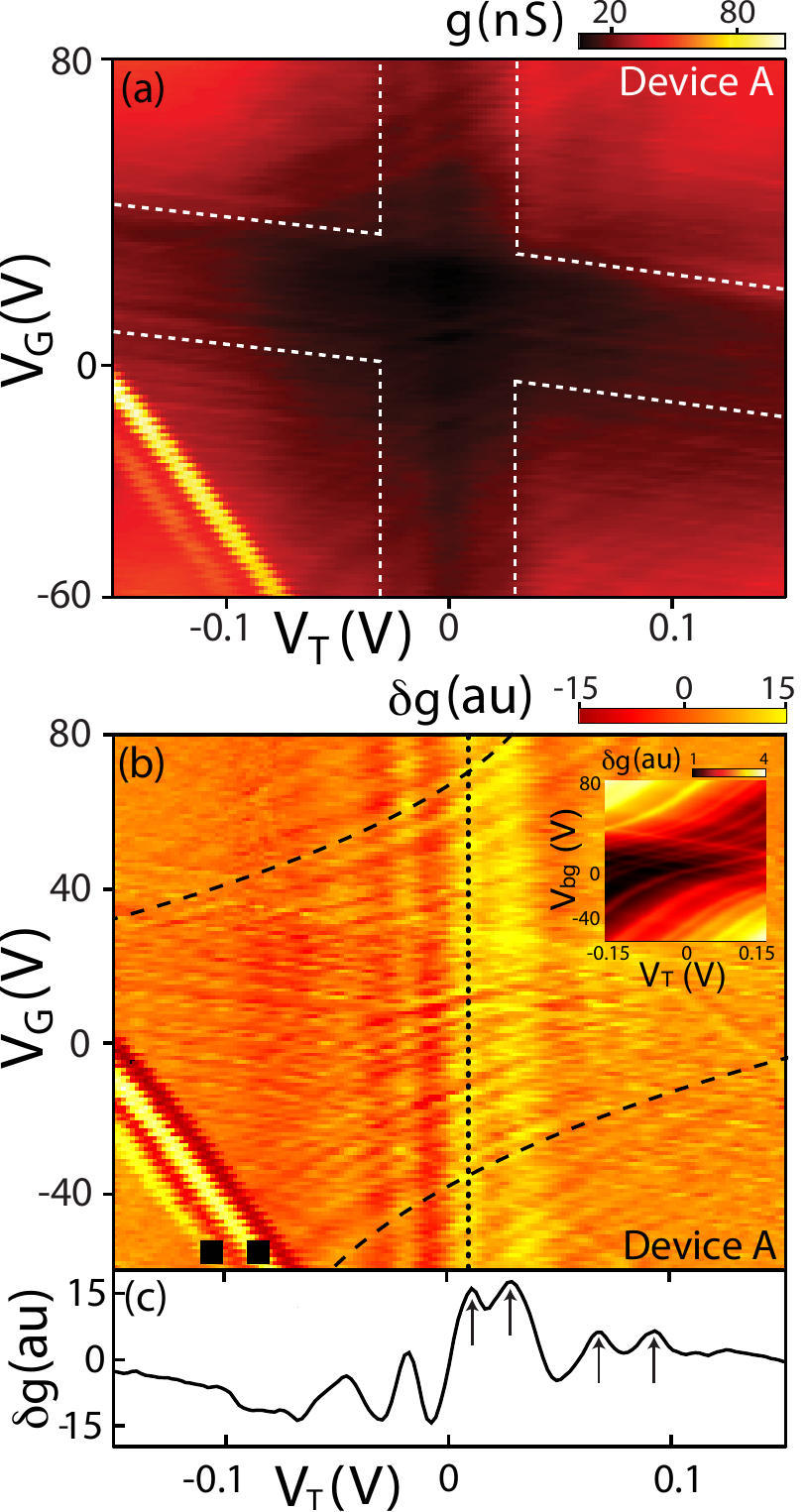}
\caption{\footnotesize{(a) $\gt(\Vt,\Vbg)$ shows a suppression of $\gt$ for the values of $\Vt$ and $\Vbg$ outlined by white dashed lines. (b) (inset) Simulation of $\dgt(\Vt,\Vbg)$. (main) $\dgt(\Vt,\Vbg)$ measured for device A.  Parabolic curves (black dashed lines) correspond to constant-density-of-states contours. In addition, a collection of $\Vbg$-independent peaks are observed (one is highlighted by the dotted line). Squares indicate two unexplained features in $\dgt$ that are likely due to tunneling enhanced by an impurity adjacent to the graphene sheet. (c) $\dgt$ averaged over all positive values of $\Vbg$ show the $\Vbg$-independent peaks apparent in (a) and (b). Arrows correspond to the voltages 10, 32, 68, 90\,mV, from left to right.}}
\end{figure}

To observe the evolution of fine features in $\gt$, a derivative with respect to the $\Vt$-axis is taken, resulting in $\dgt \propto \partial^{2} \It /  \partial{^2} \Vt$ [Fig. 2(b)]. We observe three collections of features in $\dgt$ for this range of voltages: peaks moving along parabolas in $\Vt$-$\Vbg$ space away from the CNP (dashed curves), $\Vbg$-independent peaks around $\Vt$=0 (dotted line) and strong features at negative values of $\Vt$ and $\Vbg$ (squares). $\gt$ for sample A is well approximated by $\rho(E_{\mathrm{F}}-e\Vt)$~\cite{SuppInfo}, therefore parabolic features in $\dgt(\Vbg,\Vt)$ correspond to peaks in the density of states and follow contours of constant $E_{\mathrm{F}}-e\Vt$ . These features can be fit to determine $v_{F}$ (see~\cite{SuppInfo} for information about the fit), obtaining $9.45\times10^{5}\pm8.5\times10^{4}$\,m/s, in good agreement with the theoretically expected value of $1.1\times 10^{6}$\,m/s~\cite{Brandt88}. Two very strong diagonal peaks were obtained for $\Vt$ between $\sim$-0.74\,V and -0.15\,V in the range of $\Vbg$ between -60\,V and 0\,V (squares). These two peaks only appear in this device and are not symmetric with respect to the CNP or $\Vt$. While the origin of these two features is not known, it is likely that these are associated with a resonance in an impurity adjacent to the sheet of graphene. A simulation of the tunnel conductance in the presence of disorder is shown in inset of Fig. 2(b), reproducing the features indicated by dashed curves of Fig. 2(b) but not the features labeled by the dotted line or squares (see~\cite{SuppInfo} for details of the simulation). The $\Vbg$-independent peaks, highlighted by the dotted line, are most clearly seen in Fig. 2(c) where a cut of $\dgt(\Vt)$ that has been averaged over all positive values of $\Vbg$ is shown. Four peaks are identified by arrows at 10, 32, 68, and 90\,mV (accurate to within 1\,mV), with the former two peaks appearing stronger than the latter two.

\begin{figure}
\center \label{fig3}
\includegraphics[width=3.2 in]{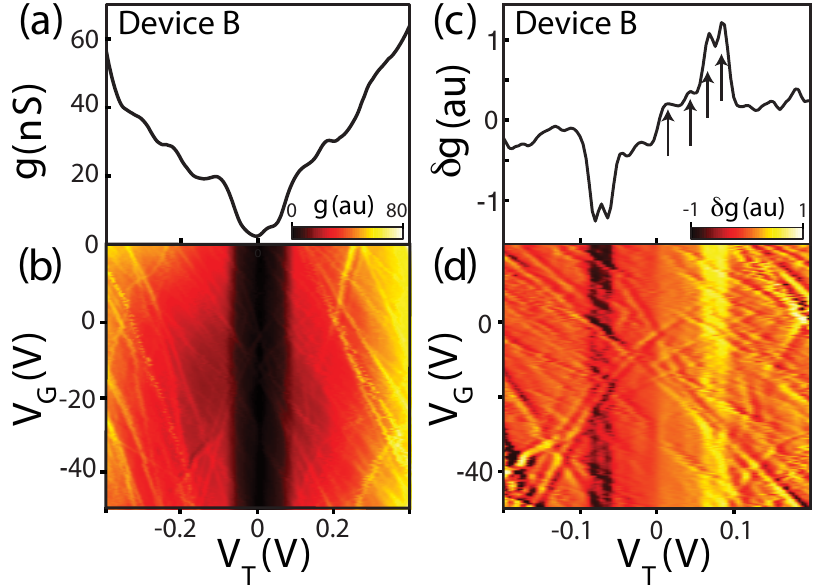}
\caption{\footnotesize{(a) Cut of $\gt(\Vt)$ at $\Vbg$=-10V for device B. showing a suppression of  $\gt$ for values of $\Vt$ less than $\sim$70\,mV. (b) $\gt(\Vt,\Vbg)$ shows the threshold of 70\,mV is independent of $\Vbg$ for the range -50\,V to 20\,V. (c) Inelastic tunneling spectroscopy, measured by $\dgt$ and averaged over the entire $\Vbg$ range, shows peaks at 10, 40, 66, 84\,mV (shown by arrows for positive $\Vt$ but also apparent for negative values of $\Vt$). (d) Peaks in $\dgt$ are $\Vbg$-independent for the range -50\,V to 20\,V.}}
\end{figure}

Similar measurements were performed on a second device, B,  where $\gt$ primarily depends on $\Vt$ and varies only slowly with $\Vbg$, a result of the tunnel transmission varying faster than the density of states as a function of $\Vt$. In Fig. 3(a), $\gt$ is suppressed for $\vert\Vt\vert<$ 70\,mV and a $\Vbg$ dependence characteristic of graphene's density of states is only observed outside this suppression. The values of $\Vt$ where this suppression occured are independent of $\Vbg$ for the range explored here [Fig. 3(b)].  A plot of $\dgt$ for device B [Fig. 3(c)] reveals peaks at similar voltages to those of device A, indicated by arrows at values of 10, 40, 68 and 84\,mV ($\pm$1\,mV), but with different intensities than device A. These peaks, like for device A, are independent of $\Vbg$ [Fig. 3(d)].

\begin{figure}
\center \label{fig4}
\includegraphics[width=3 in]{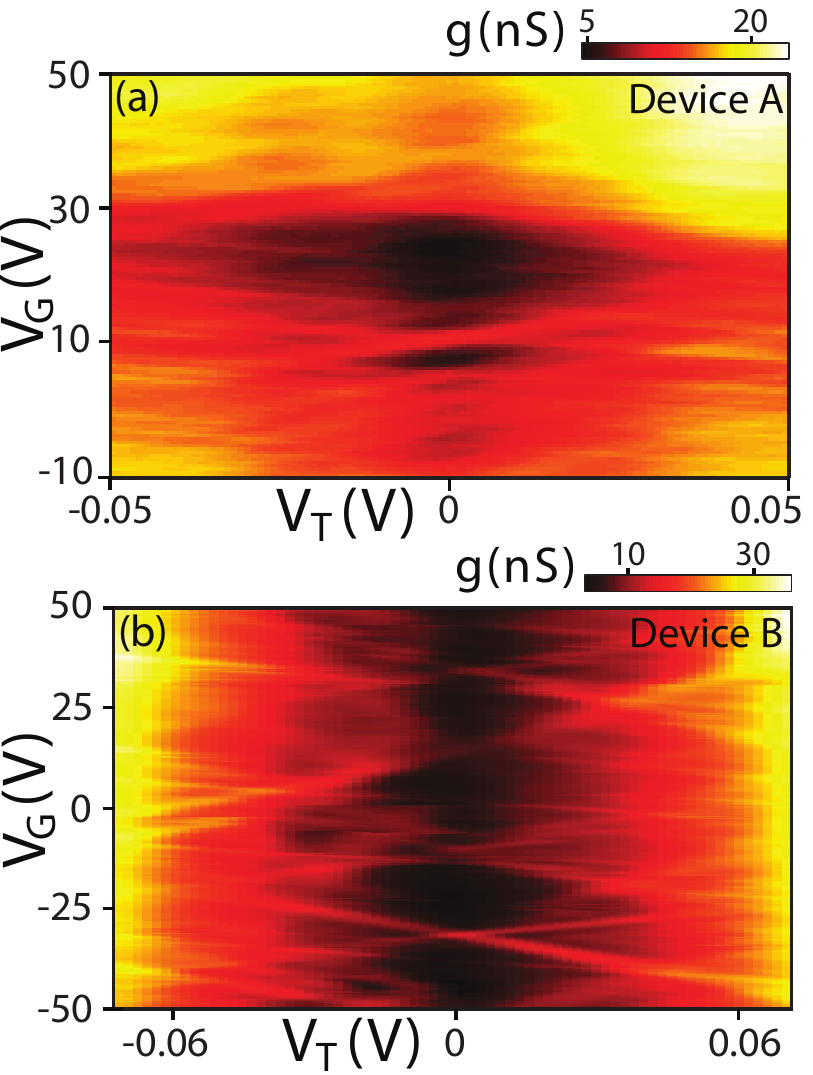}
\caption{\footnotesize{$\gt$ for smaller values of $\Vt$ for (a) device A and (b) device B. Diamond-like features are attributed to charge puddle formations in the graphene sheet for each device. Coulomb blockade features are more pronounced in device B.}}
\end{figure}

Features that are independent of $\den$ and appear as peaks in $\partial^{2} \It /  \partial{^2} \Vt$ can only reflect the variations of the tunnel transmission and not the density of states, and are ascribed to phonon thresholds enhancing tunneling. One such threshold, at 68\,mV for device A and B, has been previously measured and assigned to excitations of K-point phonons in graphene~\cite{Zhang08, Brar07}. The three other peaks are previously unreported in graphene tunneling experiments.  The lower inelastic thresholds of 10 and 40\,mV do not correspond to any phonon energy in graphite~\cite{Mohr} and are likely scattering of tunneling electrons from phonons in h-BN. The phonon spectrum in h-BN exhibits a very flat ZA branch at 40\,meV along the MK direction~\cite{Taniguchi07}. This should be associated with a Van-Hove singularity (VHS) in the phonon density of states at 40meV, explaining why an inelastic threshold is observed at this energy. The 10\,mV threshold is attributed to the ZA phonon at the A point~\cite{Taniguchi07}, where a band flattening is also observed. The highest voltage kink  (at 84\,mV and 90\,mV respectively on device A and B) is close in energy to both the M-point phonon in graphene~\cite{Mohr} and another VHS in the h-BN phonon density of states~\cite{Taniguchi07}. We also observe that peaks associated with h-BN phonons are much stronger for device A, whereas in device B, the graphene phonon thresholds are the strongest. 

The behavior of $\gt$ at low $\Vt$ provides information on the role of impurities in the tunnel transmission (Fig. 4). Both devices exhibit  Coulomb diamonds. Similar features have been found in STM experiments~\cite{Zhang09, Jung11} and are due to the formation of charge puddles around charged impurities. These puddles behave like leaky quantum dots and dominate transport in graphene at low density~\cite{DasSarma11}. For device A, these Coulomb diamonds have different sizes, indicating that several charge puddles are involved. The number and size of these diamonds increase closer to the CNP, which is due to weaker screening of charged impurities~\cite{Zhang09}. The periodicity of the conductance oscillations as a function of $\Vbg$ for device A is used to extract the average capacitance of the dots to the back gate, resulting in a typical dot size $\sim$ 200\,nm$^{2}$ (see~\cite{SuppInfo} for details of the dot size extraction), in good agreement with values reported in Ref.~\cite{Zhang09}. For device B [Fig. 4(b)], the Coulomb blockade features are much more pronounced, indicating that only a few charge puddles are involved. A charging energy of 6\,meV and a capacitance to the back-gate of $8.2\times10^{-21}$\,F are measured, which corresponds to a dot size of approximately 70\,nm$^{2}$~\cite{SuppInfo}, smaller than that of device A. The ratio $C_{\mathrm{T}}/C_{\mathrm{G}}$ for device B is found to be $\sim$100~\cite{SuppInfo}, larger than what was found for device A,  which indicates that the tunnel gate is closer to the graphene sheet in device B. Given the exponential dependence of the tunnel current on the barrier thickness, one would expect the average value of $\gt$ to be much higher for device B since the capacitance ratio is larger. This is not the case. This difference cannot be accounted for by the top gate geometry either, as the area of the top-gates for the two devices are comparable. This suggests that the tunneling area is much smaller for device B, indicating that tunneling occurs primarily at a very localized point or collection of points. This localized tunneling can result from an impurity trapped between the h-BN and the graphene, locally enhancing the tunneling rate. Another scenario resulting in localized tunneling comes from a small imperfection in the h-BN lattice, causing the tunnel barrier to be lower in one small area.

The data from device A and B show that the relative strength of elastic and inelastic tunneling strongly depends on the geometry of the tunnel junction since only B exhibits a strong suppression of $\gt$ at small values of $\Vt$ [Figs. 3(a, b)]. To our knowledge, the phonon-enhanced tunneling effect has only been observed in STM experiments where the tips were prepared to be atomically sharp (see the supplementary infomation of Ref.~\cite{Zhang08}). This suggests that inelastic scattering plays a more important role when the tip wavefunction is spatially localized, and therefore has a very broad momentum distribution. h-BN enables us to observe tunneling across a two-dimensional interface: the tunneling electrons have a well-defined parallel momentum and inelastic tunneling is suppressed.

This work was supported by the Center on Functional Engineered Nano Architectonics (FENA), the W. M. Keck fondation and the Stanford Center for Probing the Nanoscale (CPN)  . We thank H. C. Manoharan for valuable discussions.

\end{document}


\title{Supplementary information for \\``Tunneling Spectroscopy of Graphene-Boron Nitride Heterostructures"}
\author{F. Amet}
\affiliation{Department of Applied Physics, Stanford University, Stanford, CA 94305, USA}
\author{J. R. Williams}
\affiliation{Department of Physics, Stanford University, Stanford, CA 94305, USA}
\author{A. G. F. Garcia}
\affiliation{Department of Physics, Stanford University, Stanford, CA 94305, USA}
\author{M. Yankowitz}
\affiliation{Department of Physics, Stanford University, Stanford, CA 94305, USA}
\author{K.Watanabe}
\affiliation{Advanced Materials Laboratory, National Institute for Materials Science, 1-1 Namiki, Tsukuba, 305-0044, Japan}
\author{T.Taniguchi}
\affiliation{Advanced Materials Laboratory, National Institute for Materials Science, 1-1 Namiki, Tsukuba, 305-0044, Japan}
\author{D. Goldhaber-Gordon}
\affiliation{Department of Physics, Stanford University, Stanford, CA 94305, USA}

\date{\today}
\maketitle

\section{Device geometry}

The graphene flakes are annealed in Ar/H$_2$ at 350\,$^o$C and boron nitride flakes are transferred on top of them prior to any other processing, which allows for the interface between the two flakes to be very clean (See Fig. S1 for an optical image of a completed device).  The top-gated part of the graphene flake is several square microns large. However, the tunnel conductance is an exponential function of the barrier thickness, so the effective tunneling area depends strongly on the cleanliness of the interface. In fact, and as speculated in the main article, impurities in between the boron nitride and graphene can alter the nature of tunneling, resulting in the differences observed in the tunnel conductance $g$ of device A and B.  

The measured capacitance ratio $C_{T}/C_{G}\approx 72$(100) for device A(B) is smaller than the theoretical value of 150 given by a parallel plates model for this geometry. This difference is not fully understood but can be due to an effective dielectric constant lower than expected for the h-BN layer, or to an imperfect screening of the electric field of the gates by the graphene sheet.

\section{Charge puddle size}

Near the charge neutrality point in graphene, the density of carriers breaks up into a series of n-and p-type puddles~\cite{Martin08, Zhang09}, which behave as quantum dots~\cite{Jung11}. The typical size of the charge puddles in our devices can be extracted from the Coulomb diamonds observed in Fig. 4 of the main article. If one defines the capacitance ratios $\alpha_{i} = C_{i}/C_{total}$ where $i$ refers to the top gate ($i$=T) and back gate ($i$=G) capacitances, the edges of the Coulomb diamonds have slopes given by:

\begin{equation}
\frac{\partial V_{G}}{\partial V_{T}}=\frac{1-\alpha_{T}}{2\alpha_{G}} \mbox{ and }\frac{-1-\alpha_{T}}{2\alpha_{G}}
\end{equation}

From fits to the edges of the diamonds, it is therefore possible to extract the capacitance ratios $C_{T}/C_{G}$ of 72 and 100 respectively for device A and B.

The capacitance of the puddle-induced quantum dots to the back gate is given by the periodicity of conductance oscillations at $V_{T}=0$ and as a function of the back gate voltage. We substract a smooth background from $g(V_{T}=0,V_{G})$ and calculate the fast Fourier transform of $\delta g$. For example in the case of device A, we find a periodicity $\Delta V_{G}$ of 7\,V, which corresponds to a capacitance to the back gate of approximately $2\times 10^{-20}$\,F. The back gate capacitance per unit area of the silicon oxide layer has been measured to be $\approx$ 12\,nF/cm$^{-2}$ on different devices, which allows for an estimation of the typical dot size: $200$\,nm$^{2}$ in the case of device A, similar to that observed in Ref.~\cite{Zhang09}. A similar procedure is used to extract a puddle area of $72$\,nm$^{2}$ for device B.

\section{Extraction of the Fermi velocity}

The tunnel current can be expressed as:

\begin{equation}
I(V_{T}) \propto \int^{0}_{-eV_{T}}\rho(E_{F}+\epsilon)T(\epsilon,eV_{T})d\epsilon.
\end{equation}

It follows that $g$ is given by:

\begin{eqnarray}
g_{t}=\frac{dI}{dV_{T}}\propto e\rho(E_{F}-eV_{T})T(-eV_{T},eV_{T}) \\ \nonumber
+  \int^{0}_{-eV_{T}}\frac{d}{dV_{T}}\rho(E_{F}+\epsilon)T(\epsilon,eV_{T})d\epsilon
\end{eqnarray}

Using the WKB approximation it is possible to estimate the tunnel transmission as a function of the barrier thickness $d$, the transverse electron mass in boron nitride $m$, the average barrier height $U$ and the parallel momentum of the tunneling electron $k_{//}$:

\begin{equation}
T(-eV_{T},eV_{T})= exp\left(-\frac{2d\sqrt{2m}}{\hbar}\sqrt{U+\frac{(\hbar.k_{//})^{2}}{2m}-\frac{eV_{T}}{2}}\,\right)
\end{equation}

In our case, the barrier height U is comparable to half of the band-gap in boron nitride which we approximate by 4\,eV. Moreover, for electrons tunneling elastically at the K point, the parallel momentum $k_{//}$ is approximately equal to $K\approx1.7$\,$\AA$ which is fairly high. As a consequence the tunneling energy $eV_{t}$ is small compared to the effective barrier height: $U+\frac{(\hbar.k_{//})^{2}}{2m}$, and the tunnel transmission $T(-eV_{T},eV_{T})$ varies slowly with $V_{T}$ as long as inelastic tunneling is negligible, as observed on device A~\footnote{This approximation breaks down if inelastic tunneling can't be neglected, as observed on device B. In that case, K out-of-plane phonons have been shown to considerably lower the effective barrier height, and as a consequence, the tunnel transmission varies much faster as a function of $V_{t}$.}.
In that case, as a first approximation, one can neglect the variations of the tunnel transmission and write:

\begin{eqnarray}
g=\frac{dI}{dV_{t}}\propto \rho(E_{F}-eV_{T})+ \frac{dE_{F}}{d(eV_{T})} \int^{0}_{-eV_{T}}\frac{d}{d\epsilon}\rho(E_{F}+\epsilon)d\epsilon\nonumber
\\
\propto \rho(E_{F}-eV_{T}) + \frac{dE_{F}}{d(eV_{T})}(\rho(E_{F})-\rho(E_{F}-eV_{T}))
\end{eqnarray}

In graphene the Fermi energy is proportional to $\sqrt{n}$ and the derivative $ \frac{dE_{F}}{d(eV_{T})}$ should diverge close to the charge neutrality point. However, we've seen that the density of states saturates at a constant value at low carrier density [see Fig.~1(a) of the main article], and this divergence does not occur. As a consequence, the second term of Eq. 5 remains very small compared to $\rho(E_{F}-eV)$ for all applied voltages used in the experiment and the contours of constant tunnel conductance are very well approximated by curves of constant $E_{F}-eV$.

When the carrier density in the graphene sheet is large compared to the intrinsic doping $n_{0}$,  the Fermi energy is given by: 
\begin{equation}
E_{F}=\hbar v_{F}\sqrt{\pi n}=\hbar v_{F}\sqrt{\frac{\pi}{e}(C_{T}V_{T}+C_{G}V_{G}+en_{0})}
\end{equation}

Using this expression it is possible to find an analytical expression for the contours of constant $E_{F}-eV$. We find that these are parabolas of constant curvature:
 \begin{equation}
\frac{\partial^{2}V_{G}}{\partial V_{T}^{2}}\approx\pm \frac{2e^{3}}{\pi C_{G}(\hbar v_{f})^{2}}.
\end{equation}

Knowing the back gate capacitance, we can estimate the Fermi velocity from fits to these parabolas, and find a value of $9.45\times 10^{5}$\,m/s.

\section{Simulation of the tunnel conductance}

Our goal in the simulation was not to determine the density of states for disordered graphene from ab-initio calculations, but to see what are the contours of constant tunnel conductance for a density of states resembling what was observed in Fig. 2 of the main article. To this end, we estimated the tunnel conductance from the WKB formula with an empirical density of states reproducing the main features in Fig. 2 (main article). We neglected variations in the tunneling transmission, assuming that $g\propto \rho_{G}(E_{F}-eV_{t})$. The Fermi energy itself is calculated by integration of the density of states. 

The density of states  is approximated by the theoretical expression: $\rho_{G}(E)\propto \vert E\vert$ which is smoothly truncated to $\rho_{G}(E)=\rho_{0}$ under a cutoff energy that we take equal to 0.1\,eV. Randomly placed Lorentzian peaks of random widths and heights are added to this expression to simulate the resonant peaks we observed in Fig. 1(a) (main article). An example of the density of states we use is displayed on Fig. S2.
We then numerically integrate this density of states to get the Fermi energy as a function of the carrier density, and calculate the tunnel conductance $g_{t}$ as a function of both gate voltages (Fig. S3). We see that resonant peaks in the density of states give rise to two sets of curves: diagonal straight lines corresponding to constant $E_{F}$ lines, and curves of constant $E_{F}-eV$, similar to the observed features of Fig. 2(b) (main article).

\vspace{100mm}

\begin{figure*}[b!]
\center
\includegraphics[width=6 in]{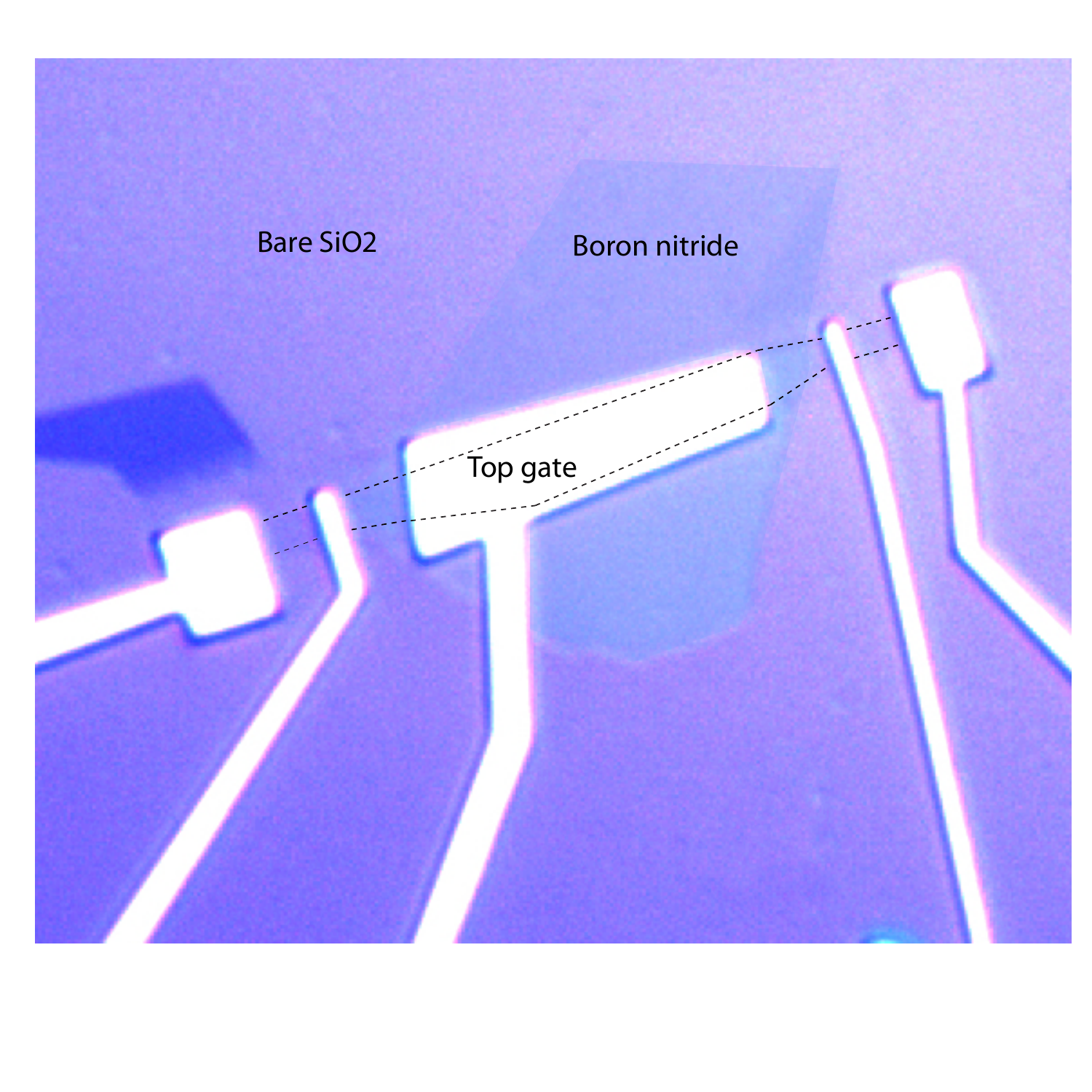}
\label{fig1}
\caption{\large{Graphene tunnel junction with a 2\,nm thick h-BN tunnel barrier. The contacts and top gate are made by standard e-beam lithography (Ti/Au 10nm/50nm). The edges of the graphene flake are overlaid with dashed lines for clarity}}
\end{figure*}

\begin{figure*}[b!]
\center \label{fig3}
\includegraphics[width=6 in]{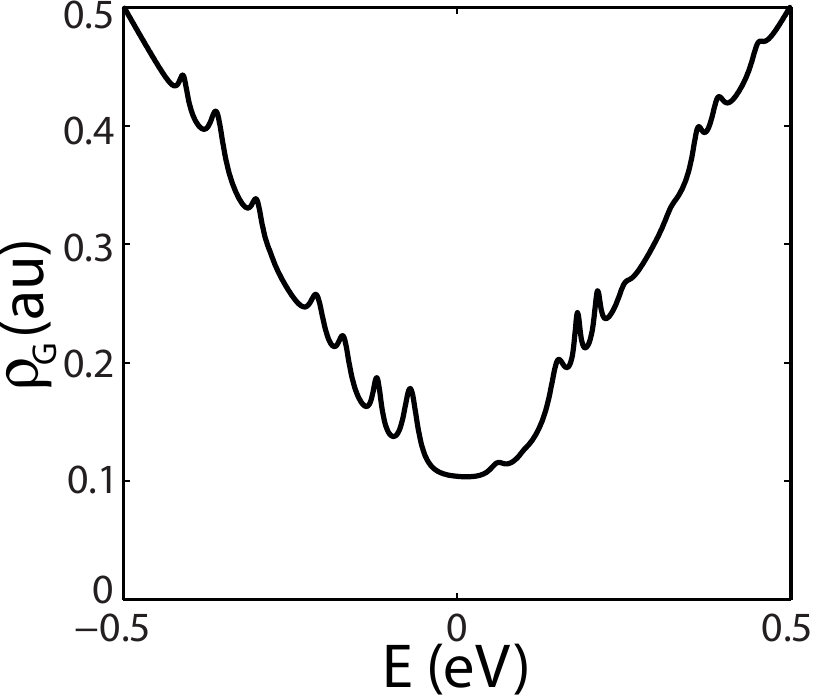}
\caption{\large{Simulation of the density of states in disordered graphene}}
\end{figure*}

\begin{figure*}[b!]
\center \label{fig4}
\includegraphics[width=6 in]{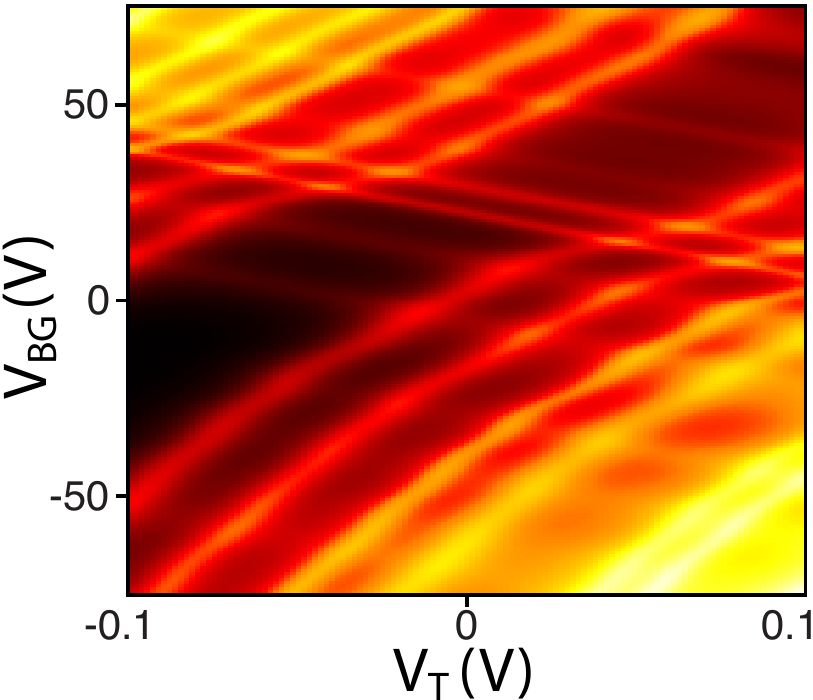}
\caption{\large{Simulation of $G_{T}$ as a function of the top gate voltage $V_{T}$ and the back gate voltage $V_{bg}$}}
\end{figure*}